# Upper limits for undetected trace species in the stratosphere of Titan[*]

Conor A. Nixon,[**a,b] Richard K. Achterberg,[a,b] Nicholas A. Teanby,[c] Patrick G. J. Irwin,[c] Jean-Marie Flaud,[d] Isabelle Kleiner,[d] Alix Dehayem-Kamadjeu,[d,e] Linda R. Brown,[f] Robert L. Sams,[g] Bruno Bézard,[h] Athena Coustenis,[h] Todd M. Ansty,[i] Andrei Mamoutkine,[j] Sandrine Vinatier,[b] Gordon L. Bjoraker,[b] Donald E. Jennings,[b] Paul. N. Romani,[b] F. Michael Flasar.[b]

In this paper we describe a first quantitative search for several molecules in Titan's stratosphere in Cassini CIRS infrared spectra. These are: ammonia ($NH_3$), methanol ($CH_3OH$), formaldehyde ($H_2CO$), and acetonitrile ($CH_3CN$), all of which are predicted by photochemical models but only the last of which observed, and not in the infrared. We find non-detections in all cases, but derive upper limits on the abundances from low-noise observations at 25°S and 75°N. Comparing these constraints to model predictions, we conclude that CIRS is highly unlikely to see $NH_3$ or $CH_3OH$ emissions. However, $CH_3CN$ and $H_2CO$ are closer to CIRS detectability, and we suggest ways in which the sensitivity threshold may be lowered towards this goal.

## 1 Introduction

The atmosphere of Titan, Saturn's largest moon, exhibits the greatest diversity of chemicals found in any planetary atmosphere outside of the Earth's. The first of these – $CH_4$, $H_2$, and $C_2H_2$ - were detected through spectroscopy using ground-based telescopes,[1,2,3] an arduous task requiring removal of telluric absorption. The first 'golden age' of multiple gas discoveries however came with the Voyager 1 flyby in 1980, when many new species were detected in the stratosphere using the IRIS infrared spectrometer, including HCN, $C_2H_4$ and $C_2H_6$ [4]; $HC_3N$, $C_2N_2$, $C_4H_2$ [5]; $C_3H_4$ and $C_3H_8$ [6]; and $CO_2$ [7]. $N_2$ was infered from ultraviolet spectroscopy. [8] Later CO and $CH_3CN$ were added to the roster using ground-based telescopes, [9,10] while $H_2O$ and $C_6H_6$ were found in the stratosphere using the Infrared Space Observatory (ISO). [11,12]

The second 'golden age' of gas discovery on Titan came with the arrival of the Cassini Saturn orbiter in 2004, but this time the detections did not flow from infrared spectroscopy of the middle atmosphere, or indeed from any remote sensing technique. Rather, it was in the ionosphere where the Cassini mass spectrometer (INMS) showed a veritable bounty of molecular species occupying all niches in the mass spectrum to the detection limit at 100 amu; [13] while the CAPS plasma spectrometer showed heavy negative ions extending to many thousands of amu beyond. [14] Detailed modelling of the INMS cracking patterns based on open-source ion (OSI) spectra [15] has proposed the presence of species unseen in the stratosphere such as ammonia ($NH_3$), toluene ($C_7H_8$), tri- and tetra-cetylene ($C_6H_2$, $C_8H_2$), methylenimine ($CH_2NH$), and several nitriles including vinyl cyanide (acrylonitrile, $C_2H_3CN$) and ethyl cyanide (propionitrile, $C_2H_5CN$). However, a later evaluation of the closed-source neutral (CSN) mass spectra, which are less contaminated by wall reactions, found confirmation only for ammonia on the above list - beyond the species already known from the infrared. [16]

Most published models [17-20] of Titan's chemistry have focused on neutral chemistry in the upper and middle atmosphere - prior to the surprise Cassini revelations of the complex ions. These invoked radical species to initiate the chemical cycle, derived from the precursor molecules $N_2$, $CH_4$ (bulk atmospheric constituents) and $H_2O$ (from external flux) by either solar photolysis or magnetospheric electron impacts. These basic radicals, including $CH_3$, $CH_2$, CH, N, H and OH, are the building blocks of the more complex species: e.g. $CH_3$ + $CH_3$ → $C_2H_6$. The models necessarily include many species not yet observed in Titan's neutral atmosphere: some are likely unstable intermediaries, e.g. $C_4H_6$, but others are stable and yet remain undetected, such as $C_3H_6$ and $C_2H_3CN$.

In this paper we report on a preliminary search for four of these predicted species in Titan's stratosphere by analysis of infrared spectra from the Cassini Composite Infrared Spectrometer (CIRS). These are the gases ammonia ($NH_3$), formaldehyde ($H_2CO$), methanol (CH$_3$OH) and the previously-discovered CH$_3$CN (acetonitrile), which has so far eluded detection in the infrared portion of the spectrum. An infrared route to measurement would permit efficient global and temporal mapping, which cannot be achieved by either in situ mass spectrometry, or ground-based sub-millimeter observations that do not resolve Titan's disk.

---

** Fax: (+1) 301 286-0212; Tel: (+1) 301 286-6757; E-mail: conor.a.nixon@nasa.gov
a *Department of Astronomy, University of Maryland, College Park, MD 20742, USA.*
b *Planetary Systems Laboratory, NASA Goddard Space Flight Center, Greenbelt, MD 20771, USA.*
c *Atmospheric, Oceanic and Planetary Physics, University of Oxford, Oxford, OX1 3PU, UK.*
d *LISA, CNRS, UMR 7583, Universités Paris 7 et Paris Est, 61 Avenue Général de Gaulle, 94010 Créteil, France.*
e *Department of Physics, College of Biological and Physical Sciences, University of Nairobi, P.O. Box 30197, Nairobi, Kenya.*
f *Jet Propulsion Laboratory, California Institute of Technology, Pasadena, CA 91109, USA.*
g *Pacific Northwest National Laboratory, P.O. Box 999, Richland, WA 99352, USA.*
h *LESIA, Observatoire de Paris-Meudon, 92195 Meudon Principal Cedex, France.*
i *Department of Space Science, Cornell University, Ithaca, NY 14853, USA.*
j *Adnet Systems, Inc., Rockville, MD 20852, USA.*

In the next section we introduce the CIRS instrument and describe the observations, followed by the data analysis method and then the results. This is followed by a discussion, and then conclusions, including the needed future work to expand this study.

## 2  Experimental

### 2.1 The Cassini Composite Infrared Spectrometer

CIRS is a dual Fourier-Transform Spectrometer (FTS), sensitive to a far-infrared spectral range (10-600 cm$^{-1}$, 1000-17 µm) using a polarization-splitting (Martin-Puplett) interferometer, and to a mid-infrared range (600-1400 cm$^{-1}$, 17-7 µm) using a conventional amplitude splitting (Michelson-type) interferometer. Both interferometers share a common 50 cm diameter telescope, fore-optics, scan mechanism and reference laser, and can achieve identical spectral resolutions ranging from 0.5 to 15 cm$^{-1}$, depending on the commanded scan time (mirror displacement). In the far-infrared, a single bolometer (known as focal plane 1, or FP1) detects the incident radiation, having a relatively large, circular field-of-view with a Gaussian response of Full Width to Half Maximum (FWHM) 2.5 mrad. The mid-infrared focal plane in contrast consists of two linear 1x10 HgCdTe pixel arrays sensitive to different sub-ranges: focal plane 3 (FP3) is operational from 600-1100 cm$^{-1}$ (17-9 µm), whilst focal plane 4 (FP4) detects the shortest wavelengths from 1000-1400 cm$^{-1}$ (9-7 µm). The CIRS focal planes are depicted in Fig. 1. In this paper, only spectra from the mid-infrared range are considered (FP3 and FP4). Further information regarding the instrument can be found in the literature. [21,22]

### 2.2 Observations

For this study, we expect to be limited in our sensitivity to the weak species by the random noise level in the spectrum, which defines the smallest emission that we can detect at the 1-σ level (one standard deviation). A 3-σ signal is normally considered the minimum to claim a detection. As our instrumental noise should in general be uncorrelated (White or Gaussian noise), we expect that by co-adding spectra we can reduce the noise level considerably. I.e., if we define the noise equivalent spectral radiance (NESR) on a single spectrum as $\sigma_1(v)$, then the spectral noise on a set of $N$ spectra should be reduced to $\sigma_n(v)=\sigma_1(v)/\sqrt{N}$. This assumes the absence of non-random effects such as instrumental artefacts, which are in fact a significant limitation at some wavelengths.

Therefore, we chose for our upper limits study to use several recent observations designed for the purpose of maximising the signal-to-noise ratio (SNR) of trace atmospheric species as much as possible, by simultaneously increasing the signal and decreasing the noise as much as possible. The increased signal is attained by pointing the instrument at Titan's limb, i.e. along a path that does not intersect the surface, while the decreased noise is attained by co-adding large numbers of spectra.

Limb observations by CIRS have previously been used to detect very weak isotopic variants of major gases, including H$^{13}$CN, HC$^{15}$N, $^{13}$CH$_3$D, $^{13}$CO$_2$, CO$^{18}$O, H$^{13}$CCCN and C$_2$HD. [23-27] However, these prior detections used relatively small spectral sets recorded at a single latitude, or were obliged to average across wide ranges of latitude to reduce the noise level, because there are typically few spectra recorded by CIRS at identical altitudes, latitudes and times. This follows from the second major advantage of limb-sounding compared to nadir (surface-intersecting) viewing, which is the ability to make resolved measurements over a wide vertical range. This arises because the opacity for trace species is usually less than unity, and therefore the radiance contribution arises almost entirely from the localized atmospheric region at the tangent altitude of the limb ray. CIRS performs around one dozen standard types of Titan observations, designed for various specific purposes such as measuring vertical aerosol opacity variation in the far-infrared, or global mapping of temperatures in the mid-infrared.

For limb measurements of gas vertical abundance profiles, the usual paradigm is to place the mid-infrared detector arrays perpendicular to the disk edge (i.e. radially from Titan's centre) to enable simultaneous measurements over a range of altitudes by virtue of the 10-element detector arrays. These observations (known as MIRLMBINTs), while excellent for vertical mapping of the more abundant trace gases, are not however ideally suited for new detections, as the spectra from different altitudes must be modelled separately and cannot be co-added to produce a single very low noise spectrum. Moreover, the atmosphere thins with altitude and the temperature changes, therefore for most minor gases the absolute signal peaks in the lower stratosphere (depending on latitude) and declines below and above.

In this work, we analyse observations of a new type (known as MIRLMPAIRs) that were custom-designed for the purpose of enabling new molecular detections – or putting low limits on non-detected species. In these observations (see Fig. 2) the CIRS mid-infrared arrays are placed parallel to the limb, with the advantage that the spectra from all ten detectors in each array fall at nearly the same altitude and within a narrow range of latitudes (depending on distance from Titan) and can therefore be cautiously co-added into a single long-path and low-noise spectrum. The caution arises from the need to consider lateral variations in gases and temperature, which are small at equatorial latitudes but change quickly at mid and high northern latitudes, at the current season. CIRS has only five electronic channels for each ten detectors, therefore an onboard pairing process is used to coadd adjacent pixels before the spectrum is returned.

Two such MIRLMPAIR sequences were performed, centred on 25°S (flyby T55) and 76°N (flyby T64) latitude, with the lower FP3 arrray positioned near 110 km (8 mbar) and 225 km (0.3 mbar) respectively. A drawback of this type of design is that, compared to the radial mode, the FP3 and FP4 detector arrays no longer cover the same range of altitude. This means the temperature sensing that is usually achieved from the methane $\nu_4$ band at 1304 cm$^{-1}$ on FP4 no longer corresponds to the right altitude to give modelling input for the FP3 spectrum, which is targeted lower. Therefore in order to model the FP3 spectra, additional temperature measurements must be used. For the T55 sequence we derived lower-stratosphere temperatures from a nadir mapping sequence (MIDIRTMAP), while for the T64 sequence we used a vertical limb measurement (MIRLMBINT, Fig. 3) that covered the full range of needed altitudes. See Table 1 for details.

# 3 Model

### 3.1 Atmospheric model

Our Titan initial model atmosphere was similar to that used in our previous study of Titan's propane bands, [28] with 100 layers equally spaced in log pressure from 1.45 bar (0 km) to 4x10$^{-8}$ bar (about 700 km), and is shown in Fig. 4. The *T-p* (temperature-pressure) profile is derived from the Huygens HASI results, [29] with altitudes computed from hydrostatic equilibrium. The initial abundance profiles of major known gases are mostly assumed to be uniformly mixed with altitude in the stratosphere, at values appropriate for Titan's equatorial regions. [30,31] The exception is acetylene, which has a vertical stratospheric gradient at low latitudes. [32] All the gases with the exception of the diatoms, $H_2$ and $N_2$, and ethylene ($C_2H_4$), condense near the tropopause and their abundance drops to low values. Isotopologues of these gases were also included as needed, i.e. $^{13}CH_4$ and $CH_3D$. Not all known Titan species or isotopologues were included in the model (e.g. CO, $HC_3N$, $C_4H_2$), if they did not have bands near the spectral regions of interest for the sensitivity study (see §3.6).

Profiles for the four test gases were later introduced in a very simple form: uniformly mixed above the tropopause and dropping quickly to an insignificant value at 45 km. The stratospheric abundance was varied as required in the sensitivity study. The final element of the atmospheric model was a haze absorber, uniformly mixed above the tropopause but variable in particle density, and using optical properties derived from a study of laboratory tholins, [33] which allowed a good fitting to the continuum over small intervals.

### 3.2 Spectral synthesis

The emerging radiance was computed based on the model atmosphere and spectral atlases for the required gases, using the NEMESIS computer code developed at Oxford University. [34] This code has been extensively validated, being previously use to model spectra of planets including Venus, Mars, Jupiter, Saturn, Uranus, and also many studies of Titan CIRS data.

The spectral atlases used for gaseous opacity are as described in a previous paper, [28] except for the addition of the four new gas species as noted here. For $NH_3$ and $CH_3OH$ we extracted line data from the HITRAN 2008 atlas. [35] For $H_2CO$ we used a new spectral line atlas containing ~5600 transitions from 877 to 1500 cm$^{-1}$, subsetted from a more comprehensive study of the $\nu_2$, $\nu_3$, $\nu_4$ and $\nu_6$ bands, [36] to include only the spectral range overlapping with CIRS. The $\nu_6$ band ($CH_2$ wag) centred on 1167 cm$^{-1}$ is the one of interest in this study. Finally, we used a new unpublished list for the $\nu_7$ band (1041 cm$^{-1}$) of $CH_3CN$ created for CIRS use as described in the next section.

The correlated-*k* method [37,38] for fast computation of radiances was used, due to the prohibitive slowness of calculating the Voigt lineshape over substantial spectral ranges in real time for use in retrievals. Therefore we computed *k*-tables for each gas over the necessary range of temperatures and pressures of the Titan atmospheric profile, including convolution with the CIRS highest spectral resolution of 0.48 cm$^{-1}$ and using 50 *g*-ordinate sampling of the *k*-distribution.

Finally, to model each recorded CIRS spectrum, the spatial extent of the detector footprints on Titan's limb had to be considered. A means of accurately modelling the MIRLMBINT type observations has been reported, which begins with laboratory-measured spatial responses for each detector [39] and co-adds these at appropriate altitudes to arrive at an 'exact' spatial weighting function. [28] We followed this treatment, sampling the spatial weighting function at 10 km intervals (about 1/5 of a Titan atmospheric scale height in the stratosphere) to enable computation in a reasonable time, while maintaining sufficient accuracy.

However, the MIRLMPAIR observations are somewhat different, as the detectors are placed horizontally on the limb, and because the detector signals are added pair-wise on the spacecraft, so that for example detectors 1 and 2 become a virtual detector '1+2', with a spatial footprint about double the size of each individual detector (ignoring a small gap between them). In the MIRLMPAIR observations, the long direction of these detector pairs is also along the array direction (*Z*). To model these, we simply assumed that they have a boxcar response cross-section in the short direction (*X*) of 0.273 mrad, i.e. vertically. We then convolved this function scaled appropriately for distance, with a sampling of rays emerging from the limb at 10-km vertical increments as before.

### 3.3 Creation of $CH_3CN$ line atlas

The $CH_3CN$ linelist is based on a very preliminary analysis of the $\nu_7$ fundamental near 10 μm. In 2006, experimental line positions were obtained by peak finding from one high resolution room temperature FTS spectrum (see description of our previous study [40] of the 11 µm region for details). Assignments were extended and modelled up to J=56 to produce a prediction of the spectrum with intensities normalized to the published integrated intensity. [41] The resulting catalogue contained over 5300 transitions between 960 to 1135 cm$^{-1}$. The accuracies of the calculated positions ranged from 0.001 cm$^{-1}$ for assigned lines, but these deviated sharply (0.1 cm$^{-1}$) at the highest J and K, because the present model does not take into account all the interactions between the $\nu_4$, $\nu_7$ and $2\nu_8$ bands presently assigned, or with the $3\nu_8$ band, in the 10-14 µm range. [42] The accuracies of individual line intensities were not characterized, and hot band transitions were not included in the database. However, the sum of the linelist intensities (3.99x10$^{-19}$ cm molecule$^{-1}$ at 296 K) fell within 3% of the reported $\nu_7$ band strength. [41] Nitrogen-broadened half widths were set to a constant value of 0.09 cm$^{-1}$/atm at 296 K, but these were later shown to be too low, by 30% to 200%. [40]

### 3.4 Retrieval method

The fitting of the spectral data by the model is achieved by successive iterations using a non-linear least squares optimal estimation method. [43] At each iteration, a cost function is computed, similar to the weighted $\chi^2$ test between the data and model spectrum, but including an additional term to allow for an *a priori* constraint, which smooths the solution and prevents it from gravitating towards an unphysical parameter set. Model parameters (temperature and aerosol opacity, or gas abundance and aerosol opacity) are adjusted after each iteration along calculated downhill gradients to reduce the $\chi^2$ on the next iteration, until an arbitrary convergence has been reached. The retrieval formalism computes final errors on retrieved parameters, based on the initial spectral error and *a priori* error (amount of constraint). A detailed description of the method as implemented in the NEMESIS code is available in the literature. [34]

### 3.5 Upper limits method

The calculation of the upper limits proceeded in three stages. In the first stage, the $\nu_4$ band of methane from 1225-1325 cm$^{-1}$ was modelled to retrieve temperatures. NEMESIS was used to model the FP4 limb data (three of the sets described in §2.2), while the lower spectral resolution FP4 nadir data was modelled using a different correlated-*k* model. [44] The combination of these four data sets was sufficient to give temperature information at the various altitudes required to later model both FP3 and FP4 MIRLMPAIR observations. The temperature profiles were then fixed and the FP3 and FP4 MIRLMPAIR spectra were then fitted for gas and haze abundance in the spectral ranges of interest, at 9-11 µm as described in §4. Finally, the abundances of known gases and isotopes, and haze opacity were also fixed, and the new gases introduced to the profile at a range of presumed abundances. The forward model was calculated directly in a single iteration (no retrieval) to show the change to the spectrum resulting.

We follow the method used previously in a CIRS study of $C_2N_2$ detection/upper limits. [45] We first define the error weighted $\chi^2$ measure of agreement between model and data as follows:

$$\chi^2_j = \Sigma_{i=1}^{M} \frac{(I_{data}(\nu_i) - I_{model}(\nu_i, q_j))^2}{\sigma_i^2} \qquad (1)$$

where $I_{data}(\nu_i)$ and $I_{model}(\nu_i)$ are the data and model spectra respectively at wavenumber $\nu_i$, $\sigma_j$ is the random noise estimate at this wavenumber, and $q_j$ is the test abundance of the new gas that we have introduced. We also define $\chi_0^2$ as the reference case when $q_j=0$.

We sample the parameter space by trying model calculations over a wide range of $q_j$, in each case calculating the change to the $\chi^2$, defined as $\Delta\chi^2 = \chi_0^2 - \chi_j^2$. If $\Delta\chi^2$ decreases to a minimum, then a positive detection of 1-σ, 2-σ, and 3-σ is made when $\Delta\chi^2$ reaches -1, -4 or -9 respectively. Similarly, a rejection is made at the 1-σ, 2-σ, and 3-σ level if $\Delta\chi^2$ increases monotonically to reach a level of +1, +4 or +9 respectively. [46]

### 3.6 Spectral range selection

When investigating the spectra for the purpose of making gas detections or placing upper limits, several considerations affect the choice of spectral range used: (i) location of the strongest emission bands of each test gas within the CIRS spectral range; (ii) the intrinsic noise level of each of the three CIRS focal planes (FP1 highest, FP4 lowest); (iii) other features that might mask the gas signatures, including weak propane bands that are not currently modelled, [28] non-random electrical interference ('spikes') on the spectrum at known frequencies that are hard to remove, and very strong gas bands that are modellable (e.g. $C_2H_2$ $\nu_5$ and $C_2H_6$ $\nu_9$) but generally leave some residual unfitted emission above the noise level, due to slight inaccuracies in line widths and other parameters.

A good range for Titan trace gas searches using CIRS spectra is the region between 900-1150 cm$^{-1}$ (11-9 µm), where the spectrum is mostly free of other strong emission bands (the comparatively weak/sparse $\nu_7$ band of $C_2H_4$ at 949 cm$^{-1}$ is the most prominent, and models well), and the lower noise of FP3 (up to 1100 cm$^{-1}$) and FP4 (1000 cm$^{-1}$ and beyond) is available. An added bonus is the overlap in spectral range between FP3 and FP4 at 1000-1100 cm$^{-1}$, which in principle means that measurements could be made simultaneously in this spectral range at two altitudes, for the MIRLMPAIR observations. However, in practice we found that the FP3 data was noisy and suffered from some calibration problems at 1000-1100 cm$^{-1}$, therefore we exclusively used FP4 data at wavenumbers above 1000 cm$^{-1}$.

These considerations led us to select the following bands for the upper limit study: ammonia $\nu_2$ centred on 950 cm$^{-1}$; methanol $\nu_8$ (1033 cm$^{-1}$); acetonitrile $\nu_7$ (1041 cm$^{-1}$); and formaldehyde $\nu_6$ (1167 cm$^{-1}$). This leaves us with a mixture of weak and strong bands. For example, the $\nu_8$ of $CH_3OH$ is very strong, with a peak intensity of $1 \times 10^{-18}$ cm molecule$^{-1}$, [47] and excellent for our purposes; while $CH_3CN$ $\nu_7$ is much weaker, reaching an intensity of just $8 \times 10^{-20}$ cm molecule$^{-1}$. [40] In the conclusions we will address how in future studies this work might be extended to include other bands in the CIRS range, with the potential to improve the upper limits in some cases.

## 4 Results

We will not here describe the results of fitting Titan's known stratospheric gases, as the retrieval of spatial and temporal variations of temperature and gas abundance from Cassini CIRS data have been extensively published in the literature. [31-32,44,48-51] These studies used a variety of line-by-line and correlated-k forward models, including NEMESIS, and have generally very close agreement of results.

Figs. 5 (25°S) and 6 (75°N) show our results. In each case, the left-hand column shows in black the residual data spectrum in the four gas ranges after the fitting and subtraction of the emissions of known species (principally $C_2H_4$ centred on 949 cm$^{-1}$ and $CH_3D$ centred on 1156 cm$^{-1}$) . In the case of $CH_3OH$ and $CH_3CN$, we avoided the $\nu_{20}$ band of propane at 1054 cm$^{-1}$, for which currently no line atlas is available and hence we cannot model, by restricting our search to the region 1030-1050 cm$^{-1}$. The coloured lines in each case again show the residual, but after a large amount of the test gas is added to the model calculation, for the purpose of identifying the locations of the strongest features of the test gas. The right-hand column shows the curve of $\Delta\chi^2$ for each species, the change in $\chi^2$ of fit that occurs as the abundance of the test gas is varied, computed across the whole range to improve sensitivity.

In none of the eight cases was any convincing minimum seen (i.e. $\Delta\chi^2=-9$ for a 3-$\sigma$ detection), although two very slight minima at the 1-$\sigma$ confidence level occurred: for $CH_3OH$ at 25°S ($\Delta\chi^2=-1.68$, abundance $3 \times 10^{-9}$) and $CH_3CN$ at 76°N ($\Delta\chi^2=-1.82$, abundance $3 \times 10^{-7}$). As the significance is so low we infer non-detections in all cases based on the present sampling, although these indications should encourage further study. Vertical dashed lines on the charts show the abundances corresponding to the 1-$\sigma$, 2-$\sigma$, and 3-$\sigma$ upper limits respectively.

Table 2 enumerates the derived upper limits at each level of significance. The most restrictive upper limits resulted for $NH_3$, despite the higher NESR on FP3, whereas the least constraining limits resulted for $CH_3CN$ due to the much weaker spectral band used. The upper limits were also lower (better) for the 25°S spectra than those at 76°N, due to the lower altitude of this observation, resulting in a significantly higher atmospheric opacity.

## 5 Discussion

### 5.1 Ammonia

$NH_3$ is the likely source of Titan's present $N_2$-dominated atmosphere, via photolysis of out-gassed ammonia shortly after Titan's formation. In the present atmosphere ammonia could be present from either episodic outgassing, that also releases the methane needed to refuel the carbon-photochemistry, or else as an intermediate product of the chemistry that continuously processes $N_2$ and $CH_4$ into more complex substances, including nitriles (-C≡N).

The 1-D photochemical model of Wilson and Atreya (hereafter WA04) [20] shows two sources of $NH_3$ production: one in the lower stratosphere at ~120 km from cosmic-ray ionization of $N_2$ resulting in a mole fraction of $9.6 \times 10^{-13}$, and a second in the upper atmosphere due to electron recombination of ammonium ions. This second source results in a relatively high abundance in the ionosphere (~$10^{-8}$-$10^{-7}$), which rapidly decreases to lower altitudes, reaching a minimum below $10^{-14}$ near 300 km. The ionospheric source was confirmed by the Cassini INMS experiment, which found a volume mixing ratio (VMR) of $7 \times 10^{-6}$ at 1100 km using the OSI mode, [15] and a slightly higher amount of $3 \times 10^{-5}$ using the CSN recordings at a similar level. [16]

Therefore, the CIRS 3-$\sigma$ upper limits derived here: $1.3 \times 10^{-9}$ at (107 km, 25°S), and $1.4 \times 10^{-8}$ at (224 km, 75°N) do not produce any meaningful constraint on photochemical models at the present time, nor are likely to: although CIRS measurements could perhaps provide a route to constraining the size/ occurrence of ongoing outgassing events.

### 5.2 Methanol

Methanol is expected to be a very minor component of Titan's stratosphere, and occurs through the combination of the $CH_3$ and OH radicals, the latter due to the small amount of water detected in Titan's atmosphere (about $8 \times 10^{-9}$ at 400 km). [11] In the model of WA04, methanol reaches a peak value of a few $10^{-10}$ at ~700 km, and declines steadily through the stratosphere, reaching $10^{-12}$ between 200-300 km.

As in the case of ammonia, our derived 3-$\sigma$ upper limits for $CH_3OH$: $1 \times 10^{-8}$ at (247 km, 25°S), and $7 \times 10^{-8}$ at (348 km, 75°N), do not provide a strong restriction on the chemistry despite the relatively intense band available for the purpose. We note that an

upper limit of $3\times10^{-8}$ was inferred from the Cassini INMS OSI data at 1100 km. [15]

### 5.3 Acetonitrile

Following the discovery of this molecule via ground-based sub-mm spectroscopy, [10] the vertical profile was subsequently measured by the same technique (IRAM telescope) applied to the 147.6 and 220.7 GHz transitions measured at 78 kHz resolution. [52] This study revealed disk-averaged vertical profile of a few $10^{-8}$ above 150 km increasingly slowly with altitude to 500 km, much more similar to the shallow HCN profile than the steep one inferred for $HC_3N$. The 1-D model of WA04 also exhibits a relatively small vertical gradient for $CH_3CN$ in this range, although with somewhat lower abundances in the range $3-10\times10^{-9}$. ISO was unsuccessful in detecting acetonitrile at 1041 cm$^{-1}$, placing an upper limit of $5\times10^{-10}$.

Cassini CIRS measurements of the vertical profiles of HCN and $HC_3N$ by limb sounding [32,53-54] reveal a more nuanced picture, with both HCN and $HC_3N$ showing much steeper gradients in the equatorial regions and southern hemisphere, and shallow gradients in the north during the early Cassini epoch of Titan northern winter/spring. This results in large equator-to-north polar enhancements in abundance at 3 mbar, attributed to downward advection in a global Hadley cell, [55] as predicted by General Circulation Models (GCMs). [56] This large polar enhancement in nitriles should apply to acetonitrile as well, and therefore we might expect an enrichment of ~4 in the lower stratosphere as seen in HCN, a species with very similar photochemical lifetime, [20] and therefore abundances of up to $\sim10^{-7}$ at latitudes above 50°N.

Our 3-σ upper limit of $1\times10^{-7}$ (247 km, 25°S) is therefore about one order of magnitude greater than the IRAM disk-average value (usually representative of low latitudes) at the same altitude, while our limit of $1\times10^{-6}$ at (348 km, 75°N), is around two orders of magnitude greater than the IRAM value at this altitude, and does not yet provide a strong constraint on the expected enrichment factor.

Finally we note that in a pre-Cassini study, [57] the 3-σ sensitivity of CIRS to detection of various nitriles in nadir viewing was predicted. For acetonitrile, this was estimated to be $2\times10^{-8}$ at 363 cm$^{-1}$ ($\nu_8$ band) and $4\times10^{-8}$ at 717 cm$^{-1}$ ($2\nu_8$ band), some order of magnitude less than our estimate, despite the lower intensities of these bands (4.5 and 8.5 cm$^{-2}$ atm$^{-1}$ respectively) compared to the $\nu_7$ (21 cm$^{-2}$ atm$^{-1}$). [58] A likely contributing factor for this difference is the non-optimal altitude probed here by CIRS FP4, as FP3 was instead targeted at the lower stratosphere where the gas pressure (and hence opacity for optically thin species) was 30x higher.

### 5.4 Formaldehyde

Formaldehyde is produced in Titan's atmosphere via the reaction of two HCO (formyl) radicals that are produced when an H atom attaches to CO. [20] WA04 predict an abundance of $1\times10^{-9}$ at 4 mbar (130 km) in the lower stratosphere, with a declining abundance above to 400 km when the profile increases again. Note that oxygen species (CO, $CO_2$) do not appear to show significant polar enhancement, [31,54,59] and we presume the same may be true for $H_2CO$. Our 3-σ constraints of $2\times10^{-8}$ at (247 km, 25°S) and $1\times10^{-7}$ at (348 km, 75°N) are therefore not restrictive at the present time.

## 6 Conclusions and further work

This work has provided a preliminary search for $NH_3$, $CH_3OH$, $CH_3CN$ and $H_2CO$ in Titan's stratosphere from Cassini CIRS spectra, resulting in upper limits for these species at two (latitude, altitude) coordinates. $NH_3$ and $CH_3OH$ in particular are predicted to occur at very low abundances, almost certainly out of reach of CIRS detection, whilst $CH_3CN$ and $H_2CO$ are predicted at abundances within 1-2 orders of magnitude of the current upper limits, holding out the possibility of detection by CIRS if the sensitivity of the search technique can be improved. There are two main routes to this goal: (i) by utilizing different spectral regions, where the combination of stronger gas bands and/or lower spectral noise produces a more favourable situation for detection, and (ii) targeting different altitudes and/or latitudes.

In the case of $CH_3CN$, a more intense band than the $\nu_7$ currently used exists in the CIRS FP4 range: the $\nu_6$ centred at 1450 cm$^{-1}$ with an intensity nearly 6x greater. The current impediments to exploiting this band are two-fold: (i) the $\nu_6$ (1379 cm$^{-1}$) and $\nu_8$ (1468 cm$^{-1}$) bands of ethane which are clearly seen on the CIRS spectra, but for which we do not have a good line atlas, and (ii) aliasing in the CIRS spectrum which is predicted to become important at ~1430 cm$^{-1}$. [28] Both these modelling issues can in principle be resolved, which may open an improved route to $CH_3CN$ detection. In the case of $H_2CO$, no other exploitable bands exist in the CIRS range, however sensitivity may still be improved by targeting FP4 at a lower altitude (~125 km) where the predicted abundance is higher, and the atmosphere thicker.

Finally, future work must also expand the remit of this study to include other likely gas candidates in Titan's atmosphere, including allene ($CH_2CHCH_2$), an isomer of the already detected propyne ($CH_3C_2H$); the two isomers of $C_3H_6$: propylene and cyclopropane, and many possible nitriles.

**Acknowledgements**

CAN and DEJ were funded during this work by the NASA Cassini Data Analysis Program grant number NNX09AK55G. Research at the Jet Propulsion Laboratory (JPL), California Institute of Technology, was performed under contract with the National Aeronautics and Space Administration. I.K. thanks the Programme National de Planétologie (PNP, France) for their funding of the spectroscopy project.

**Table 1** Cassini CIRS observations analysed in this report.

| Flyby # | Observation Name | Start Date and Time | Duration | CIRS Focal Plane | Latitude Range (mean) | Altitude Range (mean) | Pressure Range of Sensitivity | Number of Spectra |
|---|---|---|---|---|---|---|---|---|
| | | | | *Temperature Retrievals* | | | | |
| T55 | MIDIRTMAP002 | 22-MAY-2009 11:26:41 | 8 hrs | 4 | 90°S-40°N | 120-220 km | 5.0-0.5 mbar | 6263 |
| T64 | MIRLMBINT002 | 28-DEC-2009 05:16:59 | 4 hrs | 4 | 75°N-76°N (75.5°N) | 100-500 km (50 km bins) | 3.8-0.0014 mbar | 86, 126, 131, 141, 57, 67, 78, 61 |
| | | | | *Upper Limits Calculations* | | | | |
| T55 | MIRLMPAIR002 | 22-MAY-2009 02:26:41 | 4 hrs | 3 | 25±2°S (25.5°S) | 97-122 km (107 km) | 7.6 mbar | 1213 |
| | | | | 4 | 25±2°S (25.5°S) | 225-275 km (247 km) | 0.27 mbar | 941 |
| T64 | MIRLMPAIR001 | 27-DEC-2009 15:16:59 | 4 hrs | 3 | 75.8±2.0°N (76.0°N) | 204-254 km (224 km) | 0.26 mbar | 517 |
| | | | | 4 | 75.8±2.0°N (75.7°N) | 325-375 km (348 km) | 0.018 mbar | 491 |

*a* Footnote text.

**Table 2** Calculated upper limits on abundances of undetected trace gases in Titan's stratosphere.

| Gas | Latitude | Pressure | Band | Wavenumber Range For Calculation | | $1\sigma$ NESR | Abundance Upper Limits (ppb[a]) | | |
|---|---|---|---|---|---|---|---|---|---|
| | | (mbar) | | Start (cm$^{-1}$) | End (cm$^{-1}$) | (nW cm$^{-2}$ sr$^{-1}$/cm$^{-1}$) | $1\sigma$ | $2\sigma$ | $3\sigma$ |
| $NH_3$ | 25°S | 7.60 | $\nu_2$ | 960 | 1000 | 2.34 | 0.59 | 0.88 | 1.3 |
| $CH_3OH$ | 25°S | 0.27 | $\nu_8$ | 1030 | 1050 | 2.16 | 6.4 | 8.2 | 10 |
| $CH_3CN$ | 25°S | 0.27 | $\nu_7$ | 1030 | 1050 | 1.56 | 49 | 78 | 109 |
| $H_2CO$ | 25°S | 0.27 | $\nu_6$ | 1070 | 1130 | 0.39 | 2.2 | 8 | 16 |
| $NH_3$ | 76°N | 0.26 | $\nu_2$ | 960 | 1000 | 3.47 | 2.0 | 6.4 | 14 |
| $CH_3OH$ | 76°N | 0.018 | $\nu_8$ | 1030 | 1050 | 1.19 | 42 | 56 | 72 |
| $CH_3CN$ | 76°N | 0.018 | $\nu_7$ | 1030 | 1050 | 0.86 | 660 | 830 | 1000 |
| $H_2CO$ | 76°N | 0.018 | $\nu_6$ | 1070 | 1130 | 0.27 | 17 | 63 | 130 |

[a] Parts per billion.

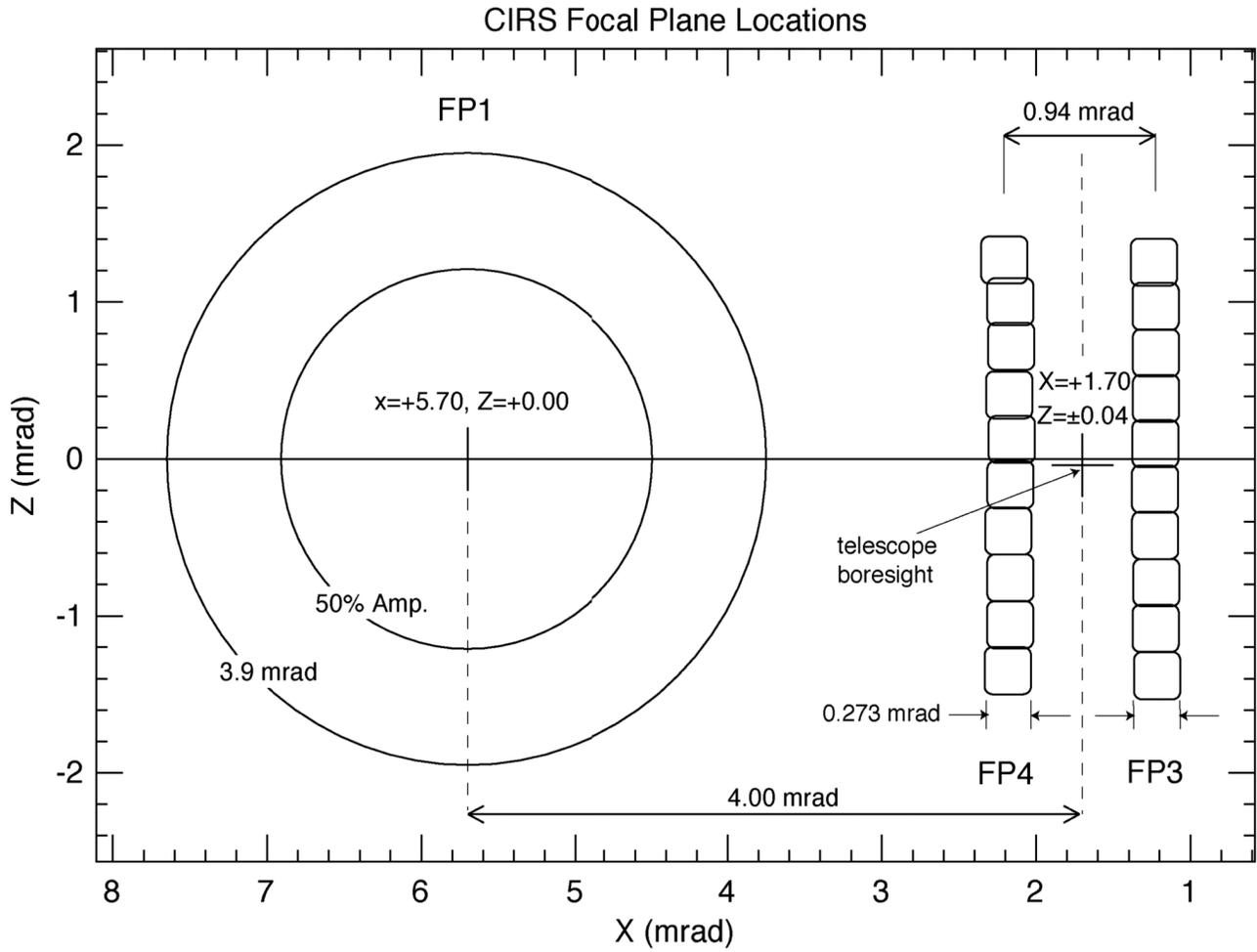

**Figure 1** CIRS instrument focal plane schematic, showing relative sizes and positions of the three detector arrays and projected footprint on sky.

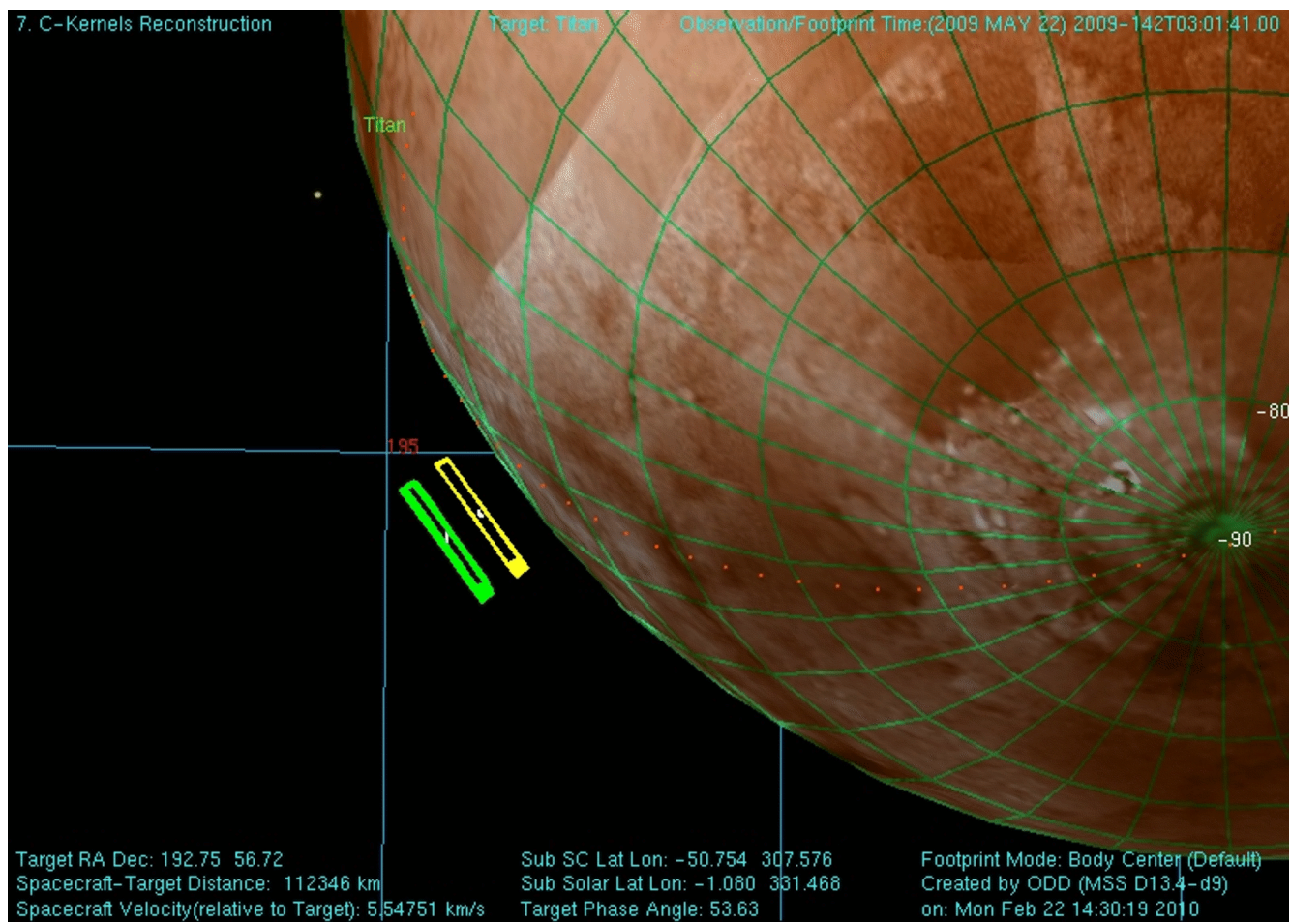

**Figure 2** Pictorial representation of the CIRS Titan mid-infrared limb pair (MIRLMPAIR) observation on the T55 flyby, May 22[nd] 2009. The FP3 array, fixed at 107 km altitude and oriented horizontally is shown in yellow and FP4 (0.94 mrad above) is in green. During the course of the observation the range from Titan increases from 109,000 to 178,000 km (113,000 depicted), and therefore the FP4 altitude steadily increases from 230 to 290 km. All detectors are used in pair mode, so the arrays are composed of effectively five oblong pixels rather than ten squares.

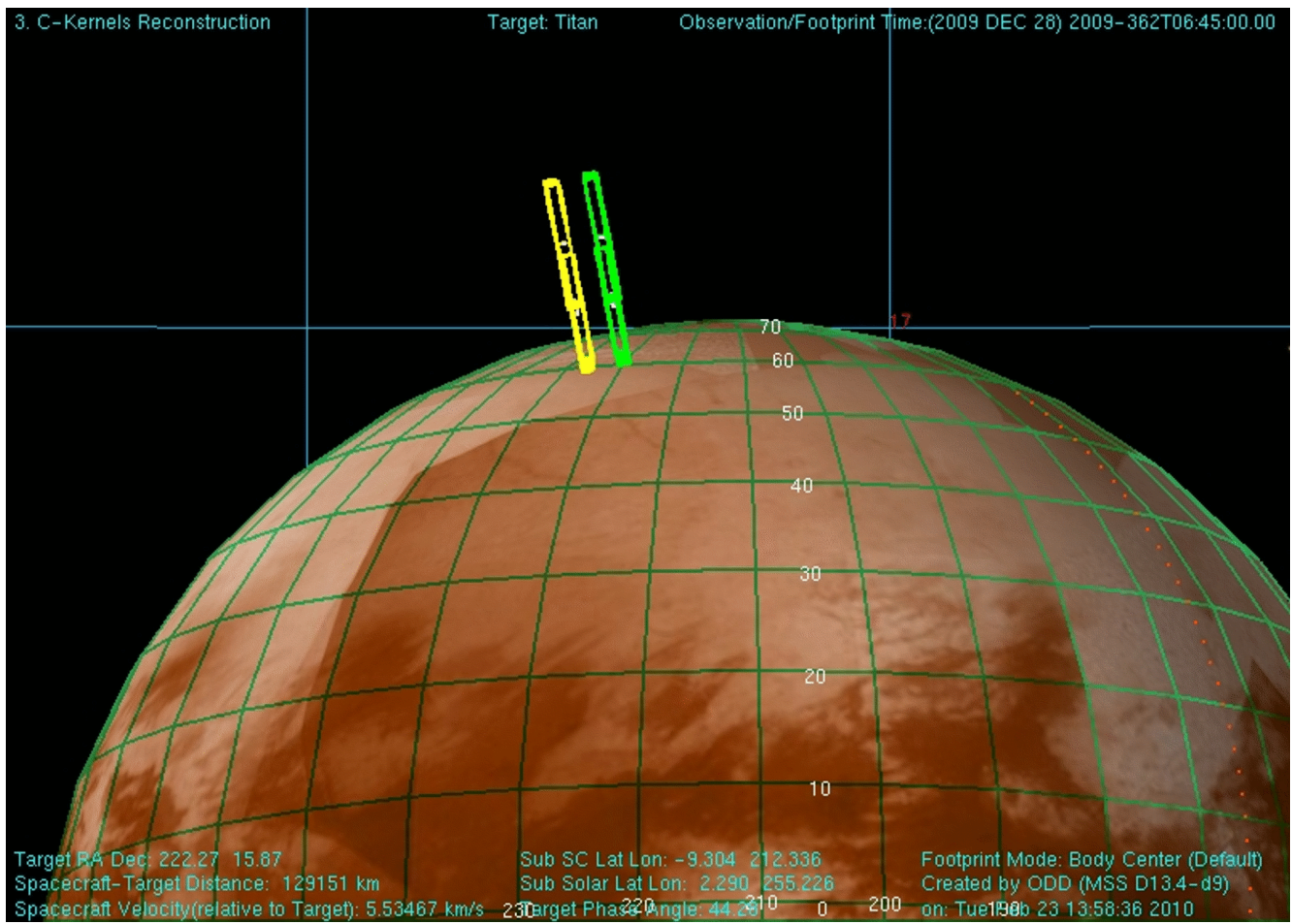

**Figure 3** Pictorial representation of the CIRS Titan mid-infrared limb integration (MIRLMBINT) observation on the T64 flyby, December 28[th] 2009. The FP3 array is shown in yellow and FP4 in green (side-by-side). Two vertical positions centred on 350 km, and ~100 km are used sequentially to measure the vertical profile of gas abundances in the atmosphere. The range is from 113,000 to 177,000 km (129,000 km depicted) and the arrays cover a vertical extent of ~400 km (each position).

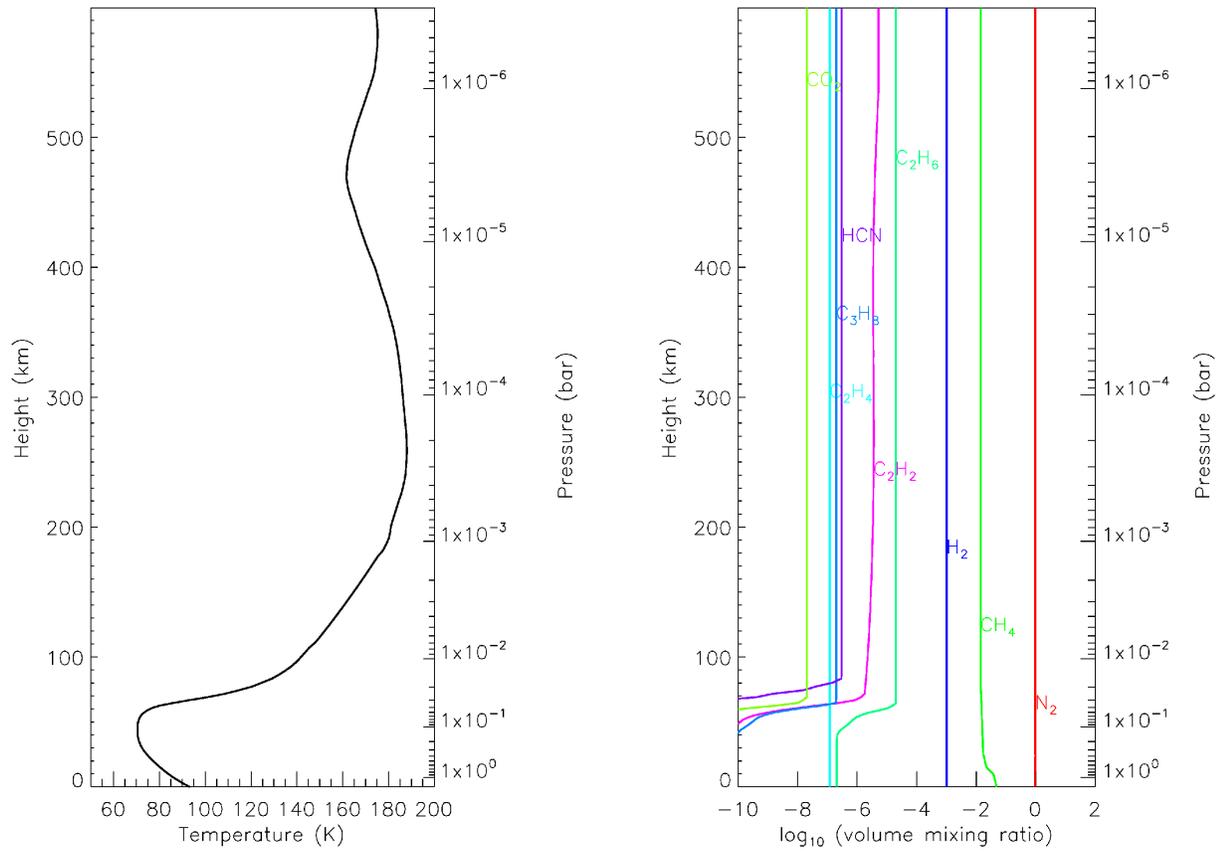

**Figure 4** Titan model atmosphere initial temperature (left) and abundance profiles of major known gases (right).

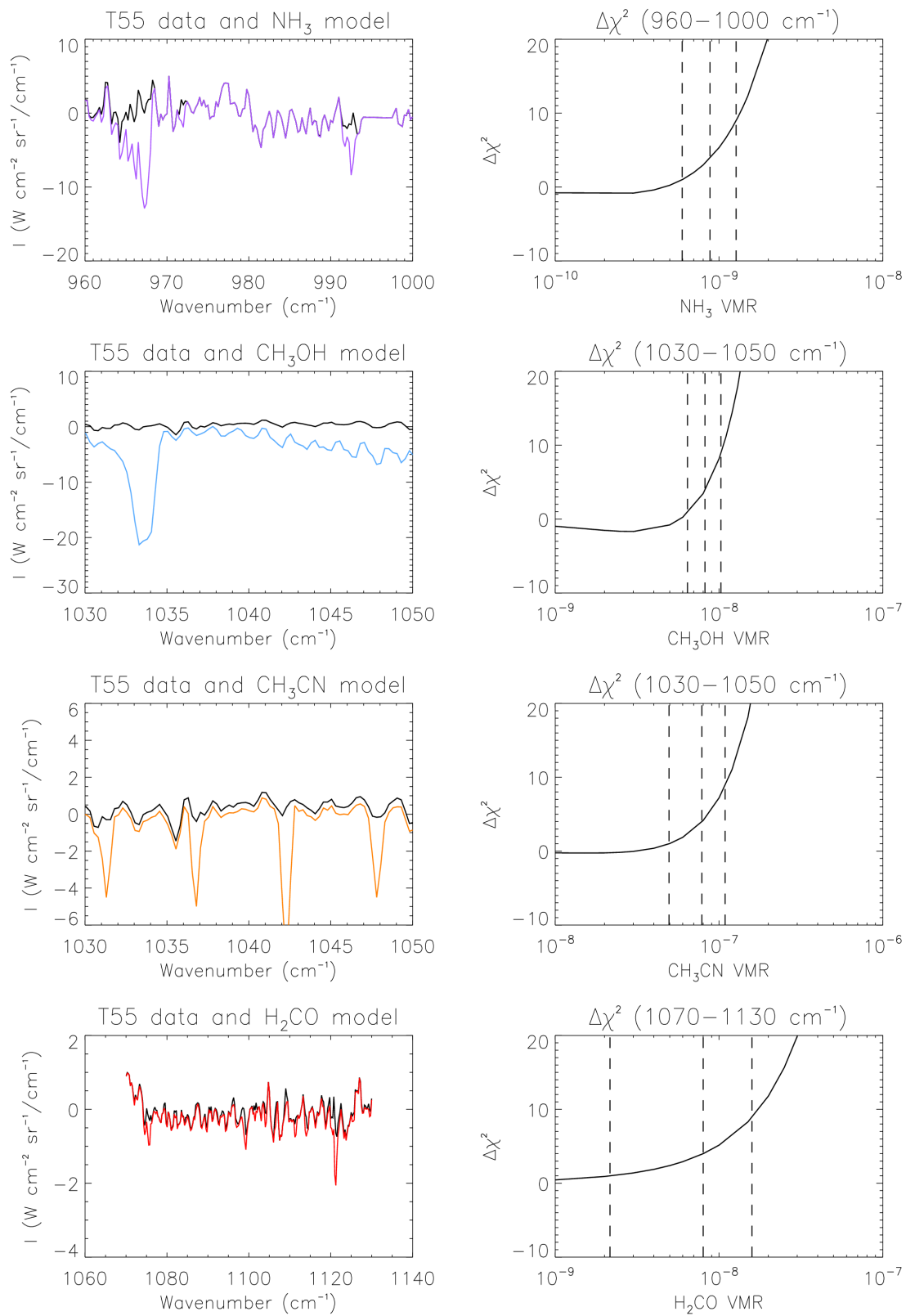

**Figure 5** Upper limits derived from T55 MIRLMPAIR data (25°S). Left column: Titan residual (data-model) spectrum in selected ranges after removal of known gases (black), and the residual found after addition of an arbitrary amount of trial gas to the model, showing location of spectral features (colored lines). Right column: the change $\Delta\chi^2$ in the goodness of fit measure $\chi^2$ plotted over a wide range of trial gas abundances (solid line). Vertical dashed lines show the abundances for which $\Delta\chi^2$ is +1, +4, +9, corresponding to gas non-detections at the 1-σ, 2-σ and 3-σ significance levels respectively.

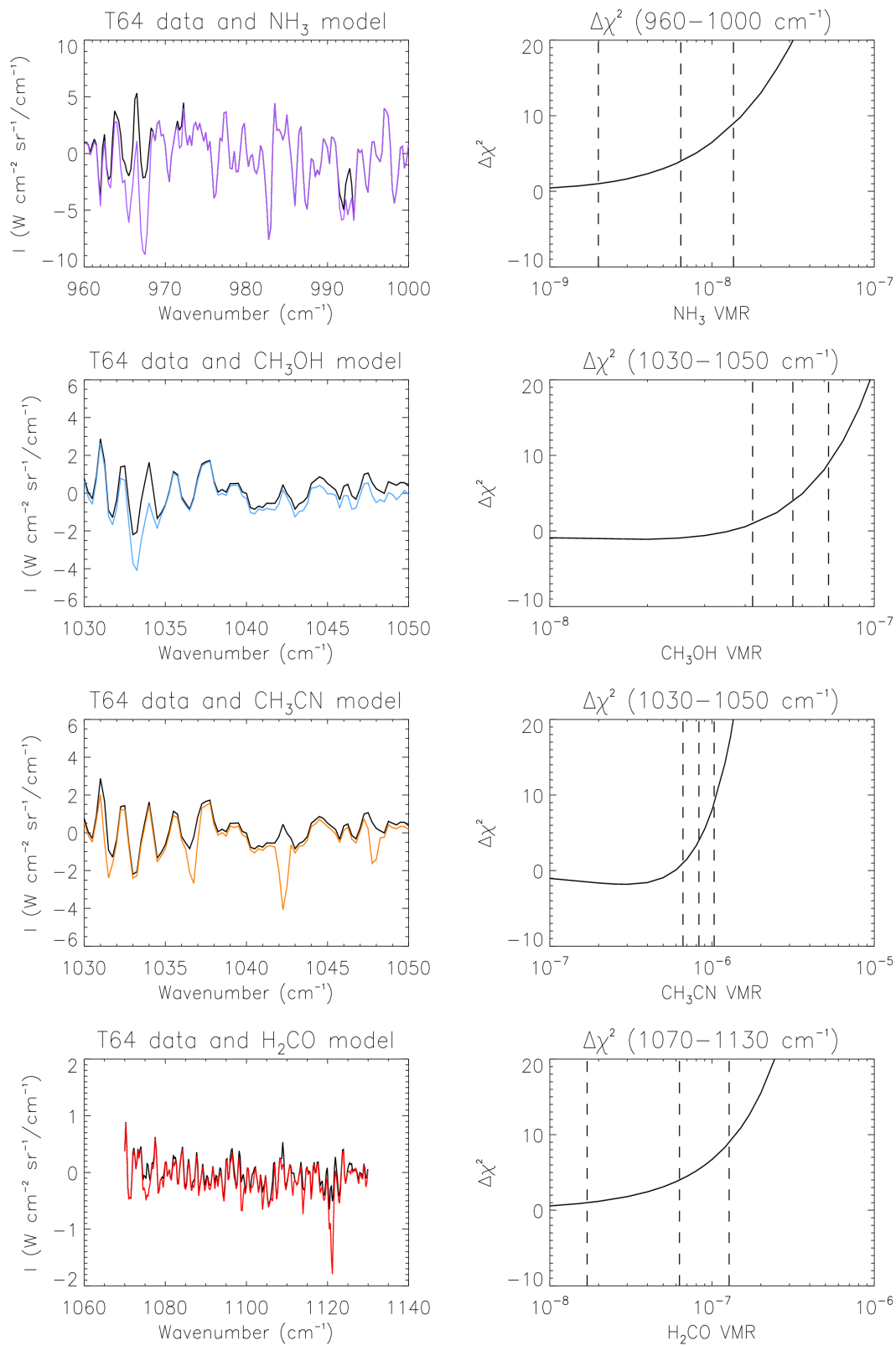

**Figure 6** Upper limits derived from T64 MIRLMPAIR data (75°N). Left column: Titan residual (data-model) spectrum in selected ranges after removal of known gases (black), and the residual found after addition of an arbitrary amount of trial gas to the model, showing location of spectral features (colored lines). Right column: the change $\Delta\chi^2$ in the goodness of fit measure $\chi^2$ plotted over a wide range of trial gas abundances (solid line). Vertical dashed lines show the abundances for which $\Delta\chi^2$ is +1, +4, +9, corresponding to gas non-detections at the 1-σ, 2-σ and 3-σ significance levels respectively.